\documentclass[acmsmall,screen,nonacm]{acmart}

\AtBeginDocument{%
  }

\graphicspath{{pics/}}
\usepackage{xcolor}
\usepackage{mdframed}
\usepackage{subcaption}
\usepackage{tabularx}
\usepackage{listings}
\usepackage{ragged2e}
\usepackage{microtype}
\usepackage{multirow}
\usepackage{enumitem}
\usepackage{siunitx}
\usepackage{adjustbox}
\usepackage{tcolorbox}
\usepackage{makecell}
\usepackage{algorithm}
\usepackage{algpseudocode}
\usepackage{graphicx}
\usepackage{booktabs}
\usepackage{array}
\usepackage{xspace}
\usepackage{wrapfig}
\usepackage{needspace}
\usepackage{booktabs}
\usepackage{graphicx}
\usepackage{multicol}

\definecolor{codebg}{HTML}{F3F5F7}
\definecolor{codeblue}{HTML}{245A8D}
\definecolor{codered}{HTML}{9B3A3A}
\definecolor{codegreen}{HTML}{3F6B4F}
\definecolor{codenumber}{HTML}{667085}
\definecolor{coderule}{HTML}{CDD3DA}

\makeatletter
\newcommand{\FinishWrap}{%
  \par
  \ifx\parshape\WF@fudgeparshape
    \ifnum\c@WF@wrappedlines>\@ne
      \@tempdima=\baselineskip
      \multiply\@tempdima by \c@WF@wrappedlines
      \vskip\@tempdima
    \fi
  \fi
  \WFclear
}
\makeatother

\lstdefinestyle{pyflowpanel}{
  language=Python,
  basicstyle=\ttfamily\scriptsize,
  keywordstyle=\color{blue!60!black}\bfseries,
  commentstyle=\color{green!30!black},
  stringstyle=\color{red!65!black},
  backgroundcolor=\color{gray!8},
  showstringspaces=false,
  keepspaces=true,
  columns=fullflexible,
  tabsize=4,
  breaklines=true,
  breakatwhitespace=true,
  numbers=none,
  frame=tb,
  rulecolor=\color{black!15},
  framerule=0.3pt,
  framesep=2pt,
  xleftmargin=2pt,
  xrightmargin=2pt,
  framexleftmargin=0pt,
  framexrightmargin=0pt,
  aboveskip=0pt,
  belowskip=0pt,
  lineskip=0pt,
  multicols={},
  float=false,
  nolol=true
}

\setlist[itemize]{leftmargin=*, nosep}
\setlist[itemize,2]{leftmargin=1em}
\setlist[itemize,3]{leftmargin=2em}

\newlength{\histtablewidth}
\newlength{\bugtablewidth}
\newcommand{\parabf}[1]{\noindent\textbf{#1}}

\newcommand{\distance}{5pt}
\newcommand{\openclaw}{OpenClaw}

\newcommand{\ourtool}{\textsc{HarnessTester}}
\newcommand{\ourbench}{\textsc{H-Bench}}

\newcommand{\agentnumber}{10}

\renewcommand\footnotetextcopyrightpermission[1]{}

\newcommand{\LDCR}{LDH\xspace}

\newcommand{\cu}{CoverUp\xspace}
\newcommand{\tta}{TestTailor\xspace}
\newcommand{\qd}{Qodo Cover\xspace}
\newcommand{\tp}{TestPilot 2\xspace}
\newcommand{\model}{GPT-5-mini\xspace}

\newcounter{finding}
\newcommand{\finding}[1]{\refstepcounter{finding}
 	\vspace{1mm}
	\begin{mdframed}[linecolor=gray,roundcorner=12pt,backgroundcolor=gray!15,linewidth=3pt,innerleftmargin=2pt, leftmargin=0cm,rightmargin=0cm,topline=false,bottomline=false,rightline = false]
		\textbf{Finding \arabic{finding}:} #1
	\end{mdframed}
	\vspace{3mm}
}

\begin{document}

\title{Complex Agents, Shallow Tests: Demystifying and Enhancing Test Adequacy of Agent Harness in the Wild}

\author{Yifan Xiong}
\affiliation{%
  \institution{University of Illinois Urbana-Champaign}
  \city{Urbana}
  \country{USA}
}
\email{yifanx14@illinois.edu}

\author{Jingyi Ge}
\affiliation{%
  \institution{University of California, Berkeley}
  \city{Berkeley}
  \country{USA}
}
\email{gjy2005@berkeley.edu}

\author{Zhenpeng Chen}
\affiliation{%
  \institution{Tsinghua University}
  \city{Beijing}
  \country{China}
}
\email{zpchen@tsinghua.edu.cn}

\author{Yiling Lou}
\affiliation{%
  \institution{University of Illinois Urbana-Champaign}
  \city{Urbana}
  \country{USA}
}
\email{yilingl@illinois.edu}

\renewcommand{\shortauthors}{Xiong et al.}

\begin{abstract}
LLM-based agentic systems are emerging as a new software paradigm. Modern agents are typically composed of backbone LLMs and a surrounding harness that serves as the operational software infrastructure for agent execution. As agent harnesses grow increasingly complex, agents suffer from diverse harness implementation bugs, raising substantial reliability concerns. In this work, we conduct the first empirical study to systematically investigate the test adequacy of harness in real-world agentic systems. Our analysis reveals that agent harness remains substantially undertested. In particular, LLM-dependent harness (LDH) code, despite its critical role in processing LLM outputs and governing agent behavior, receives limited testing attention, with less than half of its lines and branches covered by existing tests. Motivated by these findings, we further propose \ourtool{}, the first harness-oriented test generation technique that incorporates explicit agent-harness contract support to construct contract-faithful test setups and extensively exercise LDH code. Our evaluation shows that \ourtool{} substantially outperforms state-of-the-art general-purpose test generation techniques in achieving 75.95\%/84.76\% larger line/branch coverage gains and 69.89\% larger mutation-score gains. Furthermore, \ourtool{} detects 122 real-world harness bugs in widely-used agentic systems (e.g., OpenClaw), among which, 88 bugs are previously-unknown bugs and 69 bugs have been confirmed by agent developers. These results highlight the practical effectiveness of \ourtool{} in improving test adequacy and assuring the reliability of real-world agentic systems.
\end{abstract}

\maketitle

\section{Introduction}

Driven by recent advances in large language models (LLMs), LLM-based agentic systems (also called agents) are emerging as \textit{a new software paradigm} and have been increasingly adopted across a wide range of domains~\cite{claudecode, devin, cursor, codex, novikov2025alphaevolvecodingagentscientific, lu2024ai, kannan2024smart, zeng2023large}. For example, \openclaw{}~\cite{openclaw}, a recent general-purpose agent, has become one of the fastest-growing open-source software systems in history, attracting 3.2 million monthly active users, more than 346,000 GitHub stars, and over 500,000 active agent instances worldwide within just five months of its release~\cite{openclawnews, openclawnewstwo}.

An agentic system is typically viewed as the composition of  probabilistic backbone \emph{LLMs} and a surrounding \emph{harness}~\cite{li2026agent, ning2026code, anthropicharnessblog, anthropicharnessblog2, codexharnessblog}. While LLMs provide reasoning capabilities, harnesses serve as the operational software infrastructure that wraps around the LLM to provide critical functionality spanning execution environment management, context and memory management, guardrail and governance, workflow orchestration, and tool interface. Harnesses in modern agentic systems often comprise a large codebase, e.g., the harness of \openclaw{} contains over 196,000 lines of code~\cite{openclaw}. Recent studies have shown that substantial improvements in agent performance can be achieved through optimizing harnesses alone while the backbone LLMs remain unchanged~\cite{zhang2026stop, li2026agent}. Consequently, there has been increasing effort in harness engineering~\cite{zhang2026stop, li2026agent, anthropicharnessblog, codexharnessblog}, which focuses on building a more capable harness layer to unleash the capabilities of backbone LLMs. 

Despite the tremendous potential, agentic systems suffer from pervasive reliability issues~\cite{MAST, microsoftagentfailure, sun2026agentslastexam, deng2025ai, zhang2024agent, bai2026ai, xiao2025reducing}, leading to severe real-world consequences~\cite{yagoubi2026agentleak, metaagentleak, guardian2026awsoutage, aiid2025replit, cyera2026agentdamage}. Particularly, as modern agentic systems have grown increasingly complex, recent studies reveal that agent applications and frameworks suffer
from diverse harness implementation bugs~\cite{chen2026agentreactive, neuripsRahardja25, zhang2026understanding, zhu2026bugs, xue2025characterization}. These reliability concerns highlight an urgent need for systematic techniques to assure the agent reliability~\cite{MAP}.

Testing, which examines whether a software system exhibits expected runtime behaviors under given inputs, has long been a cornerstone of software reliability assurance. In particular, test adequacy criteria, such as code coverage~\cite{zhu1997software} and mutation scores~\cite{jia2011mutation}, provide important indicators of test quality by measuring how thoroughly a test suite exercises the program under test and its ability to detect bugs. To date, test adequacy has been extensively studied and widely adopted in traditional software development~\cite{weyuker1986axiomatizing,zhu1997software,jia2011mutation}.

Despite the growing adoption of agentic systems and their substantial reliability concerns, how adequately their harness implementations are tested by agent developers during development and maintenance remains largely unexplored. Although numerous benchmarks~\cite{swebench, sun2026agentslastexam, merrill2026terminal} have been proposed to evaluate agentic systems, they primarily measure task-level end-to-end success by treating agents as black boxes, without examining how thoroughly their internal harness implementations are exercised.  This leaves a fundamental question about the testing practices of real-world agentic systems: \textit{How thoroughly do agent developers test the agent harness implementations during development and maintenance, and how effective are these tests at detecting harness bugs?}

\parabf{Empirical Study.} To fill this knowledge gap, we conduct the first empirical study to systematically analyze the test adequacy of agent harness  in the wild. Our key insight is that \textit{LLM-dependent harness (\LDCR{})}, the harness code with explicit data or control dependencies on LLM outputs, deserves particular testing attention, as it is responsible for correctly processing nondeterministic LLM outputs spanning a vast value space. Recent work~\cite{chen2026agentreactive} also shows that many agent failures occur during interactions between LLMs and harness.
Therefore, beyond conventional test adequacy analysis over the entire agent codebase, we specifically investigate the test adequacy of LLM dependent harness code. To this end, we study the test adequacy of \agentnumber{} large-scale and widely-used agentic systems (e.g., \openclaw{}~\cite{openclaw-v2026-3-28}) from  diverse domains (e.g., coding, general-purpose applications, and scientific discovery), through the following research questions.

\textit{RQ1: Overall Test Adequacy.} Our quantitative analysis shows that real-world agent harness remains substantially undertested. Particularly, LLM-dependent harness code, despite its critical role in processing LLM outputs and governing agent behavior, receives limited testing attention: less than half of \LDCR{} lines and branches are covered, with an average mutation score of only 33.60\%. Such a gap leaves reliability risks from LLM-harness interactions largely unexposed by current testing practice in agent systems. Moreover, we find that achieving high \LDCR{} test adequacy is challenging, as it requires contract-faithful test setups that satisfy the expected data structures, runtime states, and interaction protocols across harness components. 

\textit{RQ2: Semantic Test Adequacy.} Beyond the aggregate coverage reported in RQ1, we characterize the semantics of \LDCR{} predicates and examine test coverage across semantic categories. Our manual analysis reveals diverse semantic patterns in what \LDCR{} predicates control, how they make decisions, and what forms of LLM-derived values they process. Predicates with explicit schemas, bounded value spaces, or directly controllable decisions (e.g., tool action and policy guards) are generally better covered, while larger gaps arise for those involving dynamic execution contexts or open-ended value spaces (e.g., agent orchestration). These results highlight the challenge of adequately testing \LDCR{} as the required test input spaces become implicit, dynamic, and difficult to enumerate.

\parabf{Harness-Oriented Test Generation Technique.}
Given the prevalent test inadequacy of agent harnesses revealed in our study, we further propose \ourtool{}, an LLM-driven test generation technique specifically designed to extensively exercise LLM-dependent harness code in agentic systems. Inspired by our finding that reaching target harness requires test setups that satisfy interaction requirements across harness components, \ourtool{} enables harness-oriented test generation with explicit support for \emph{agent harness contracts}, which specify the expected data structures, execution preconditions, interaction protocols across harness components. 

\textit{RQ3: Test Adequacy Improvement.}
Our evaluation results show that the tests generated by \ourtool{} achieve substantially higher coverage and mutation scores than all the four state-of-the-art LLM-based test generation techniques~\cite{pizzorno2025coverup, zhou2026testtailor, qodo-cover-artifact, testpilot2-artifact}. \ourtool{} outperforms the strongest baseline selected for each agent and metric by achieving 75.95\%/84.76\% larger mean \LDCR{} line/branch coverage gains and a 69.89\% larger mean mutation-score gain. Our ablation study further demonstrates the substantial contribution of agent-harness contract support in \ourtool{}. 

\textit{RQ4: Real-World Harness Bug Detection.}
We further evaluate the test adequacy of tests generated by \ourtool{} through their capability to detect real-world harness bugs. In RQ4.a, among 250 historical harness bugs, tests generated by \ourtool{} detect substantially more bugs than the best baseline (43 versus 2). In RQ4.b, we further apply \ourtool{} to the latest versions of the studied agents to discover previously unknown bugs. \ourtool{} identifies 122 harness bugs, among them, 88 are previously unknown bugs and 69 have been confirmed by developers by the paper submission time. These results demonstrate the practical effectiveness of \ourtool{} in uncovering real-world harness bugs and improving the reliability of agentic systems.

In summary, this paper makes the following contributions:
\begin{itemize}[leftmargin=5pt]
\item \textbf{Study}. We conduct the first empirical study of test adequacy in real-world agent harnesses, with a novel focus on LLM-dependent harness code. Our study not only reveals prevalent test inadequacy in agent harness but also identifies key challenges in constructing high-adequacy tests for \LDCR{}.

\item \textbf{Approach}. We propose the first harness-oriented test generation technique, \ourtool{}, which incorporates harness contract support to construct contract-faithful test execution contexts and exercise diverse LLM-dependent harness code. 

\item \textbf{Practical Impact}. Our study reveals important test-adequacy gaps and challenges in current agent testing practices. Moreover, our extensive evaluation demonstrates the practical effectiveness of \ourtool{} in improving agent reliability, which not only substantially improves the test adequacy of existing agent tests but also detects harness bugs across real-world agentic systems.

\end{itemize}
\section{Background and Related Work} \label{sec:background}

\parabf{Agentic Systems and Agent Harness.}
Modern agentic systems (e.g., OpenClaw~\cite{openclaw}) are becoming increasingly complex and are typically viewed as the composition of probabilistic backbone LLMs and a surrounding harness~\cite{li2026agent, ning2026code, anthropicharnessblog, anthropicharnessblog2, codexharnessblog}. The harness surrounding the backbone LLMs serves as the operational software infrastructure that provides critical functionality, including execution environment management, context and memory management, guardrails and governance, workflow orchestration, and tool interfaces. Meanwhile, growing research efforts have focused on harness engineering~\cite{zhang2026stop, li2026agent, anthropicharnessblog, codexharnessblog}, aiming to develop systematic methodologies for optimizing the agent harness layer. In particular, proper harness design and implementation have been shown to substantially improve the capabilities of agentic systems even when the backbone LLMs remain unchanged~\cite{zhang2026stop, li2026agent, lee2026metaharness, lou2026autoharness}. These advances highlight the importance of harness quality, aligning well with our motivation to examine the test adequacy of agent harnesses.

\parabf{Testing for Agentic Systems.}
Mainstream work focuses on benchmarking agents on end-to-end task success rates~\cite{liu2024agentbench,mialon2024gaia,zhou2024webarena,xie2024osworld,yao2025taubench,swebench}. While useful for comparing task-level capabilities, such evaluation paradigms primarily treat agents as black boxes, without examining how thoroughly their internal harness implementations are exercised. In this work, we provide a complementary perspective to existing agent benchmarks by analyzing the white-box test adequacy of harness.

Recent studies have investigated testing practices in emerging agentic systems~\cite{hasan2026testingagents,pan2026tangent,kohl2025structural,magalhaes2025untestable}. While these studies share a similar perspective with our work by studying existing developer tests in agent repositories, they mainly focus on characterizing testing patterns, such as test granularity (e.g., unit- or system-level testing). In contrast, our work focuses on the \emph{test adequacy} of agent harnesses, with a novel analysis perspective on critical LLM-dependent harness code. In addition, we further propose the first harness-oriented test generation technique, \ourtool{}, which not only substantially improves harness test adequacy but also detects real-world harness bugs.

\parabf{LLM-based Test Generation.}
A growing body of work has explored LLM-based test generation~\cite{yuan2024chattester,chen2024chatunitest,schafer2024testpilot,lemieux2023codamosa,ryan2024symprompt,wang2024hits,yang2026telpa,pizzorno2025coverup,zhou2026testtailor}, demonstrating strong effectiveness in generating diverse and semantically meaningful tests. However, existing techniques are primarily designed for general software systems. As revealed by our study, generating high-adequacy tests for \LDCR{} requires contract-faithful test setups that satisfy interaction requirements across harness components. To address this challenge, we propose the first harness-oriented test generation technique \ourtool{} that enhances LLM-based test generation with agent harness contract support. Our experiments further confirm the substantial improvements in test coverage and bug detection capabilities from \ourtool{} compared to existing general-purpose LLM-based test generators~\cite{pizzorno2025coverup, zhou2026testtailor, qodo-cover-artifact, testpilot2-artifact}.

\section{Empirical Study Setting}~\label{sec:study}
This work performs the first study to investigate the test adequacy of harness in real-world agentic systems. This section presents the setup of our empirical study. 

\subsection{LLM-Dependent Harness Code}~\label{sec:definition}
In this work, we refer to the harness code that has data dependency or control dependency on the outputs of LLM invocations as \textit{LLM-dependent harness} (\LDCR{}) code. We consider LLM-dependent harness code a critical testing target because it is responsible for correctly interpreting diverse and nondeterministic LLM outputs to govern subsequent agent execution. Errors in such code can therefore propagate through the agent workflow and lead to incorrect agent behavior.
In fact, as revealed by recent work\cite{chen2026agentreactive}, prevalent agent bugs occur between LLM-harness interactions. 

\parabf{Definition.}
Given an agentic program $A$, let $\mathcal{S}_{\mathit{LLM}} \subseteq \mathrm{Stmt}(A)$ denote the set of LLM invocation statements that produce model outputs, and let $s \rightarrow_d h$ denote that statement $h$ is directly or transitively dependent on statement $s$. We define the \emph{LLM-dependent harness code} of $A$ as
\[
\LDCR(A) =
\left\{
h \in \mathcal{H}(A)
\;\middle|\;
\exists s \in \mathcal{S}_{\mathit{LLM}}.\;
s \rightarrow_d h
\right\},
\]
where $\mathcal{H}(A) \subseteq \mathrm{Stmt}(A)$ denotes the set of harness statements. Intuitively, $\LDCR(A)$ captures harness code whose computation depends on values produced by LLM invocations. 

\parabf{Extraction.}
Existing program analysis frameworks provide general machinery for interprocedural data-flow and control-flow analysis~\cite{reps1995interprocedural,avgustinov2016ql}, but they do not directly capture the LLM-specific semantics required by \LDCR{} extraction. Particularly in agent harness, LLM outputs can originate from heterogeneous model providers and invocation abstractions, and subsequently flow through agent-specific representations, such as message objects, structured responses, and tool-call structures. Consequently, extracting \LDCR{} requires explicitly identifying LLM-output sources and modeling how output-derived values propagate through these abstractions. Therefore, to automate \LDCR{} extraction, we implement an interprocedural harness analysis tool that models LLM invocations as sources and propagates the corresponding data- and control dependency facts throughout the program until a fixed point is reached. We include all the explicit dependency through assignments, field or container accesses, parameter and return flows, and other taint-preserving transformations. Implementation details are in our artifact~\cite{harnesstester2026}. 

Listing~\ref{lst:llm-flow} illustrates an example of LLM-dependent harness code. The code highlighted in red (e.g., Line~15 and Line~18) are the sources of LLM outputs returned from the LiteLLM interface and the provider provenance of the OpenAI client; and the code highlighted (e.g., Line~29) in green are identified as LLM-dependent harness code, to which the LLM outputs propagate.  

\begin{figure}[!hbt]
\centering
\refstepcounter{lstlisting}
\label{lst:llm-flow}
\captionsetup{position=top,skip=6pt}
\caption*{\textbf{Listing~\thelstlisting.} Illustration example of LLM-dependent harness code.}

\newcommand{\sourcehl}[1]{%
  \begingroup
  \setlength{\fboxsep}{0.4pt}%
  \colorbox{red!18}{\ttfamily\scriptsize #1}%
  \endgroup
}
\newcommand{\flowhl}[1]{%
  \begingroup
  \setlength{\fboxsep}{0.4pt}%
  \colorbox{green!18}{\ttfamily\scriptsize #1}%
  \endgroup
}
\newcommand{\providerhl}[1]{%
  \textcolor{purple!75!black}{%
    \ttfamily\scriptsize\bfseries{#1}}%
}

\lstset{
  style=pyflowpanel,
  numbers=left,
  numberstyle=\tiny\color{black!60},
  stepnumber=1,
  numberblanklines=true,
  numbersep=5pt,
  xleftmargin=14pt,
  xrightmargin=2pt,
  breaklines=true,
  breakatwhitespace=false,
  multicols={},
  float=false,
  nolol=true,
  literate={},
  escapeinside={(*@}{@*)},
  postbreak={}
}

\begin{minipage}[t]{0.495\linewidth}
\vspace{0pt}
\begin{lstlisting}[firstnumber=1]
import json
from pathlib import Path
(*@\providerhl{from litellm import completion}@*)
(*@\providerhl{from openai import OpenAI}@*)
from tools import search_tool

(*@\providerhl{client}@*) = (*@\providerhl{OpenAI}@*)(timeout=30)
TOOL_REGISTRY = {"search": search_tool}

def setup_session(path: Path):
    ...
def teardown_session(path: Path):
    ...
def invoke(messages):
    return (*@\sourcehl{completion}@*)(model="gpt-5", messages=messages)

def direct_answer(prompt):
    response = (*@\sourcehl{\providerhl{client}.responses.create}@*)(input=prompt)
    return (*@\flowhl{response.output\_text}@*)
\end{lstlisting}
\end{minipage}\hfill%
\begin{minipage}[t]{0.495\linewidth}
\vspace{0pt}
\begin{lstlisting}[firstnumber=23]
def run_agent_turn(messages, session_path: Path):
    created_here = False
    if not session_path.exists():
        setup_session(session_path)
        created_here = True
    try:
        (*@\flowhl{response}@*) = (*@\flowhl{invoke}@*)(messages)
        (*@\flowhl{message}@*) = (*@\flowhl{response.choices[0].message}@*)
        results = []
        for (*@\flowhl{call}@*) in (*@\flowhl{message.tool\_calls}@*) or []:
            (*@\flowhl{tool}@*) = TOOL_REGISTRY[(*@\flowhl{call.function.name}@*)]
            (*@\flowhl{args}@*) = json.loads((*@\flowhl{call.function.arguments}@*))
            (*@\flowhl{results.append}@*)((*@\flowhl{tool}@*)(session_path=session_path, **(*@\flowhl{args}@*)))
        return (*@\flowhl{message.content}@*), (*@\flowhl{results}@*)
    finally:
        if created_here:
            teardown_session(session_path)
\end{lstlisting}
\end{minipage}

\end{figure}

\subsection{Studied Agents}
We include a set of representative open-source agentic systems based on the following inclusion criteria. (1) Literature-backed candidate pool construction: we refer to prior agent research ~\cite{majgaonkar2025understandingcodeagentbehaviour,hsu2026webds,weiss2026lacunaresearchmapmachine,yang2025rdagent,wang2026openclawsafety} for their studied agentic systems; (2) Open-source quality filtering: we only include the agentic systems that have at least 10{,}000 GitHub stars as of the data collection date and at least one official release within the preceding 18 months, and an official release or stable revision provides an executable automated test suite that can be reproduced in our environment; (3) Agent diversity: we select agents to cover different domains (e.g., software engineering, auto-research, general purpose) and two programming languages (e.g., Python or TypeScript) widely used to build agents. Table~\ref{tab:subjects} summarizes \agentnumber{} studied agentic systems with their characteristics.

\subsection{Test Adequacy Metrics}
In this study, we focus on the following test adequacy metrics.

\parabf{Coverage.} We include the widely used coverage metrics, i.e., line and branch coverage~\cite{zhu1997software}, which calculate the ratio of lines and branches exercised by tests. In particular, we collect both (i) project-wide coverage, which focuses on the coverage over the entire codebase of the studied agent repository; (ii) \LDCR-wide coverage, which focuses on the coverage over LLM-dependent harness code. We collect coverage with the widely-used coverage package~\cite{coveragepy} for Python,  and the coverage providers integrated with Vitest~\cite{vitest} for TypeScript.

\parabf{Mutation Score.}
While coverage measures how extensively code is exercised, mutation testing assesses the fault detection capabilities of tests by injecting artificial faults into the program under test. A mutant is considered as killed by a test when the test passes on the original program but fails on the mutated program; and the mutation scores calculate the  ratio of killed mutants to all the generated mutants. Therefore, to measure how sufficiently existing tests can detect the potential bugs in LLM-harness interaction, we further present the mutation scores over \LDCR{}. In particular, we apply the common mutation tools \textsc{mutmut}~\cite{mutmut} and \textsc{StrykerJS}~\cite{strykerjs} to generate mutants on LLM-dependent harness code, and present the proportion of killed mutants. 

\begin{table}[t]
\centering
\caption{Statistics of studied agentic systems.}
\label{tab:subjects}
\scriptsize
\setlength{\tabcolsep}{2.8pt}
\renewcommand{\arraystretch}{1.00}

\resizebox{0.9\columnwidth}{!}{%
\begin{tabular}{@{}llllrrrr@{}}
\toprule
\textbf{Agent} & \textbf{Lang.} & \textbf{Domain}
& \textbf{Structure} & \textbf{Version}
& \textbf{KLOC} & \textbf{Tests} & \textbf{Stars} \\
\midrule

OpenHands~\cite{openhands-1-6-0}
& Py/TS & Issue resolution & Multi-agent & 1.6.0
& 194.8 & 2,646 & 84.1K \\

Aider~\cite{aider-v0-86-2}
& Python & Interactive coding & Single-agent & 0.86.2
& 20.2 & 489 & 48.2K \\

SWE-agent~\cite{swe-agent-v1-1-0}
& Python & Issue resolution & Single-agent & 1.1.0
& 12.7 & 90 & 20.1K \\

PR-Agent~\cite{pr-agent-v0-36-0}
& Python & Code review & Single-agent & 0.36.0
& 20.9 & 478 & 12.6K \\

GPT Researcher~\cite{gpt-researcher-v3-5-0}
& Python & Research synthesis & Multi-agent & 3.5.0
& 30.7 & 47 & 29.0K \\

Browser Use~\cite{browser-use-0-13-2}
& Python & Browser automation & Single-agent & 0.13.2
& 70.9 & 869 & 109.4K \\

RD-Agent~\cite{rd-agent-v0-8-0}
& Python & Data-driven R\&D & Multi-agent & 0.8.0
& 44.5 & 75 & 14.2K \\

OpenClaw~\cite{openclaw-v2026-3-28}
& TS/JS & General-purpose & Multi-agent & 2026.3.28
& 851.0 & 2,352 & 386.4K \\

Roo Code~\cite{roo-code-v3-54-0}
& TS/JS & Interactive coding & Multi-agent & 3.54.0
& 161.7 & 549 & 24.3K \\

Kimi Code~\cite{kimi-code-0-19-2}
& TS/JS & Interactive coding & Single-agent & 0.19.2
& 166.3 & 595 & 11.2K \\

\bottomrule
\end{tabular}%
}
\end{table}
\section{RQ1: Overall Test Adequacy}
\label{sec:rq1}

This RQ performs a large-scale quantitative analysis to investigate the overall test adequacy of the existing test suite in studied agentic systems. As shown in Table~\ref{tab:rq1_ldcr_coverage},  Column ``\LDCR{} Share'' presents the ratio of \LDCR{} lines/branches to the agent codebase, Column ``\LDCR{} Coverage'' presents line/branch coverage in LLM-dependent harness code, Column ``Project coverage'' presents the project-wide line/branch coverage, and Column ``Mutation Score'' presents the mutation scores over \LDCR{}. 

\begin{table*}[!hbt]
\centering
\captionsetup{position=top,skip=3pt}
\caption{Overall test adequacy.}
\label{tab:rq1_ldcr_coverage}

\small
\setlength{\tabcolsep}{2.5pt}
\renewcommand{\arraystretch}{1.08}

\resizebox{0.75\textwidth}{!}{%
\begin{tabular}{@{}l*{7}{c}@{}}
\toprule
\multirow{2}{*}{\textbf{Agent}}
& \multicolumn{2}{c}{\textbf{\LDCR{} Share}}
& \multicolumn{2}{c}{\textbf{\LDCR{} Coverage}}
& \multicolumn{2}{c}{\textbf{Project Coverage}}
& \multirow{2}{*}[-0.6ex]{%
    \shortstack{\textbf{Mutation}\\\textbf{Score}}} \\
\cmidrule(lr){2-3}
\cmidrule(lr){4-5}
\cmidrule(lr){6-7}

& \textbf{Line}
& \textbf{Branch}
& \textbf{Line}
& \textbf{Branch}
& \textbf{Line}
& \textbf{Branch}
& \\
\midrule

OpenHands
& 16.76\% & 29.15\%
& 54.91\% & 53.42\%
& 66.65\% & 49.93\%
& 41.66\% \\

Aider
& 42.51\% & 57.57\%
& 53.87\% & 48.93\%
& 58.65\% & 49.21\%
& 35.10\% \\

SWE-agent
& 8.78\% & 12.63\%
& 44.81\% & 44.15\%
& 58.51\% & 40.86\%
& 24.11\% \\

PR-Agent
& 37.72\% & 52.30\%
& 27.09\% & 25.76\%
& 37.88\% & 27.58\%
& 16.91\% \\

GPT Researcher
& 32.43\% & 56.62\%
& 10.67\% & 11.49\%
& 25.75\% & 10.29\%
& 4.47\% \\

Browser Use
& 10.82\% & 16.79\%
& 28.30\% & 25.48\%
& 37.51\% & 25.75\%
& 25.23\% \\

RD-Agent
& 18.67\% & 27.93\%
& 5.39\% & 6.73\%
& 20.82\% & 3.61\%
& 3.58\% \\

OpenClaw
& 23.38\% & 27.45\%
& 56.73\% & 50.55\%
& 59.74\% & 51.29\%
& 53.18\% \\

Roo Code
& 17.68\% & 20.45\%
& 68.14\% & 62.18\%
& 57.33\% & 48.59\%
& 60.86\% \\

Kimi Code
& 14.72\% & 16.18\%
& 92.83\% & 83.09\%
& 71.98\% & 61.26\%
& 70.94\% \\

\midrule
\textbf{Average}
& \textbf{22.35\%} & \textbf{31.71\%}
& \textbf{44.27\%} & \textbf{41.18\%}
& \textbf{49.48\%} & \textbf{36.84\%}
& \textbf{33.60\%} \\

\bottomrule
\end{tabular}%
}
\end{table*}

\textit{\LDCR{} Prevalence.} As shown in Column ``\LDCR{} share'', \LDCR{} represents a relatively compact portion (i.e., less than 45\% of executable lines) of the agent codebase, while containing a comparatively high concentration of decisions. In particular, across all studied agents, the \LDCR{} branch share is 1.10-1.75 times its line share. One possible explanation is that LLM outputs are frequently consumed by decision-making logic, such as parsing and validating model responses, dispatching actions, and updating agent states. Meanwhile, it is also as expected that agent codebase contains substantial supporting code that does not directly depend on LLM outputs, e.g., provider initialization, model configuration, and execution environments management, which typically do not consume model-produced values. Overall, the \LDCR{} distribution confirms the necessity of further complementing project-wide test adequacy with an additional focus on LLM-dependent harness code. 

\textit{Limited Test Coverage.}
As shown in Column ``\LDCR{} Coverage'' and ``Project Coverage'', most studied agents  exhibit limited test coverage at both the project and \LDCR{} levels. Across the entire codebase, more than half of the executable lines and branches remain uncovered by existing tests on average. In addition, despite the importance of \LDCR{}, it does not consistently receive higher test coverage than the project as a whole. Specifically, \LDCR{} line coverage is lower than project-level line coverage in eight of the ten agents; for branch coverage, \LDCR{} achieves comparable or higher coverage in six agents but lower coverage in the remaining four. In summary, these results show that existing agentic systems mostly remain undertested regarding the test coverage, with a limited testing attention drawn to the critical LLM-dependent harness code.  

\textit{Limited Fault Detection Effectiveness.}
As shown in Column ``Mutation Score'', existing tests also show limited effectiveness in detecting injected mutants in LLM-dependent harness code. Mutation scores range from 3.58\% for RD-Agent to 70.94\% for Kimi Code, with an average of only 33.60\%. It indicates that, on average, about two-thirds of the injected faults in \LDCR{} cannot be detected by any existing tests. These undetected mutants arise for two main reasons: some are not even exercised by any test (e.g., 38.36\%) , while others (e.g., 25.92\%) are exercised but survive because the tests fail to propagate or capture the behavioral changes introduced by the mutations. Overall, the results further reveal the limited capabilities of existing agent tests in detecting bugs occurring in LLM-dependent harness code, implying that existing testing practices cannot fully mitigate the reliability risks arising from LLM–harness interactions during agent execution.

\finding{The harnesses of real-world agentic systems remain substantially undertested, with average line and branch coverage below 50\% and 40\% in agent codebase. More concerningly, LLM-dependent harness code, despite its high concentration of decision logic to process LLM outputs and govern agent behavior, also receives limited testing attention: existing tests only cover less than half of \LDCR{} lines and branches and exhibit limited capability to detect bugs in \LDCR{}, with only 33.60\%  mutation scores. The results reveal a substantial gap between the critical role of LLM-dependent harness code and the limited test adequacy it receives, indicating that current testing practice in real-world agentic systems cannot fully mitigate the reliability risks arising from LLM–harness interactions during agent execution.}

\textit{Agent-Wise Comparison.} 
As shown in Table~\ref{tab:rq1_ldcr_coverage}, test adequacy can vary substantially across the studied agents. For example, Kimi Code achieves relatively high project- and \LDCR{}-wide coverage (e.g., up to 92.83\% \LDCR{} line coverage) and the highest mutation score (i.e., 70.94\%) across all  studied agents. We then perform a case analysis on the test implementation in Kimi Code. 

Although all studied agents, including Kimi Code, extensively use mocks to isolate LLMs and other external components, which is a consistent observation as prior work~\cite{pan2026tangent}, simply using mocks does not necessarily make deeper harness behavior easy to reach. Model outputs often flow through multiple harness components before reaching deeper \LDCR{} logic, where each component may expect specific input structures, runtime states, and interaction sequences. Our inspection shows that Kimi Code provides systematic test support for satisfying these requirements. For example, reusable helpers construct mocked model responses with the required fields for different scenarios, such as turn completion and tool invocation. Fake LLMs and scripted generators reproduce valid interaction sequences, while local HTTP/SSE servers and environment fakes reproduce the runtime interactions required for provider communication and tool execution. Together, these utilities help tests construct valid execution contexts and vary model outputs, interactions, and runtime states to exercise different harness behaviors.
These requirements resemble the classic notion of \emph{software contracts}~\cite{meyer1992applying,beugnard1999making}, which specify the obligations and guarantees governing interactions between software components. In the context of agentic systems, these requirements form \emph{agent-harness contracts}, capturing how model-proposed actions map to tool interfaces, how returned observations correspond to the originating calls, and how subsequent decisions depend on state established through preceding interactions. Reusable test utilities in Kimi Code effectively encode many of these contracts, making it easier to construct both  contract-faithful and diverse execution contexts to exercise different decisions.

In contrast, OpenClaw, which exhibits much lower test adequacy, provides a representative counterexample. 
Although its tests also contain extensive mocks and fixtures, they are organized in a relatively ad hoc manner, making it difficult to construct contract-faithful setups across diverse test scenarios. At the model-output boundary, parsers, dispatchers, and error handlers often contain branch-rich logic over response structures and content. While a single mocked response may exercise a nominal path, exploring different decisions requires diverse responses that still satisfy the expected provider-response contracts. Reaching deeper \LDCR{} logic can be even more challenging, as model-derived values may flow through multiple components, such as agent runners and gateways. Reaching a target path therefore requires tests to configure their callbacks, exchanged data, runtime states, and environments consistently. Without reusable support for constructing such setups, developers must repeatedly identify and assemble the required test context and its variants for different execution paths. Compared with Kimi Code, OpenClaw may therefore face greater difficulty in constructing test setups that are both contract-faithful and sufficiently diverse. 

Taken together, these cases suggest that adequate \LDCR{} testing depends on test-side setups that are both sufficiently diverse and faithful to the relevant agent harness contracts. Constructing such tests, however, can be challenging in practice: agent developers are supposed to identify the component boundaries traversed by model-derived values, understand the data structures and interaction protocols expected at those boundaries, and construct mocks that collectively satisfy these contracts while supporting the diverse required behavioral variations.

\finding{Achieving high test adequacy on LLM-dependent harness code calls for contract-faithful test setups that satisfy the expected data structures, runtime states, and interaction protocols across harness components. Reusable test utilities can facilitate the construction of such valid execution contexts, while supporting the exploration of diverse inputs.}

\section{RQ2: Semantic Test Adequacy}\label{sec:rq2}

The quantitative results in RQ1 reveal overall test insufficiency in agent codebase and LLM-dependent harness code. Meanwhile, aggregate metrics alone cannot reveal whether these testing gaps are uniformly distributed in particular types of LLM-dependent harness logic. This distinction is important, as \LDCR{} code can contain rich semantics to serve different responsibilities in agent workflows, which may present different testing challenges. In this RQ, we therefore characterize the semantics of \LDCR{} branch predicates and examine test coverage across different semantic categories to understand which types of LLM-dependent harness logic are adequately covered and where important testing gaps remain. We focus on branch predicates as they often represent critical decision points that govern agent behavior and encode  rich semantics. 

To this end, we first randomly sample 700 LLM-dependent harness branch predicates from all studied agents. Specifically, we analyze each predicate from three perspectives: (i) \emph{component intention}, which refers to the responsibility that the predicate  fulfills within the agent workflow; (ii) \emph{judgment pattern}, which refers to the decision rule encoded by the predicate when evaluating LLM-derived values; and (iii) \emph{LLM value form}, which refers to the runtime representation of LLM-derived values.
As there is no existing taxonomy on semantics of \LDCR{} branch predicates, we develop our taxonomy through an open-coding process~\cite{strauss1998basics,seaman1999qualitative}. Two annotators, each with five years of software development experience and substantial experience with agents, first perform a pilot study on another small set of  predicates and assign descriptive codes without relying on predefined categories. They then discuss and consolidate the resulting codes, iteratively refining the categories and their definitions until reaching a stable codebook. Using the finalized codebook, the two annotators independently label all the sampled predicates.  Cohen's $\kappa$ on these independent annotations before resolving disagreements is 0.96, indicating high inter-annotator agreement~\cite{cohen1960coefficient}. Remaining disagreements are resolved through discussion with a third reviewer.

\begin{table*}[t]
\centering
\caption{Taxonomy and distribution of semantic patterns in \LDCR{} branch predicates.}
\label{tab:rq2_category_distribution}
\small
\setlength{\tabcolsep}{4pt}

\resizebox{\textwidth}{!}{%
\begin{tabular}{@{}c
@{\hspace{1.0em}\vrule width 0.4pt\hspace{1.0em}}
c
@{\hspace{1.0em}\vrule width 0.4pt\hspace{1.0em}}
c@{}}
\toprule

% ---------------- Component Intention ----------------
\begin{tabular}[t]{@{}lr@{}}
\multicolumn{2}{c}{\textbf{Component Intention}} \\
\cmidrule(lr){1-2}
\textbf{Category} & \textbf{Share} \\
\midrule
Provider I/O Normalization      & 29.14\% \\
Tool Action Handling            & 23.57\% \\
Context Memory Management       & 13.57\% \\
Agent Orchestration Control     & 14.00\% \\
Observation Feedback Handling   & 12.43\% \\
Error Recovery Handling         & 7.29\% \\
\end{tabular}

&

% ---------------- Judgment Pattern ----------------
\begin{tabular}[t]{@{}lr@{}}
\multicolumn{2}{c}{\textbf{Judgment Pattern}} \\
\cmidrule(lr){1-2}
\textbf{Category} & \textbf{Share} \\
\midrule
\addlinespace[1.5ex]
Presence or Shape Check   & 47.14\% \\[1.20ex]
Value or Pattern Matching & 33.29\% \\[1.20ex]
Size or Threshold Check   & 15.29\% \\[1.20ex]
Policy Guard              & 4.29\%  \\
\end{tabular}

&

% ---------------- LLM Value Form ----------------
\begin{tabular}[t]{@{}lr@{}}
\multicolumn{2}{c}{\textbf{LLM Value Form}} \\
\cmidrule(lr){1-2}
\textbf{Category} & \textbf{Share} \\
\midrule
\addlinespace[1ex]
Generated Content            & 25.00\% \\[0.45ex]
Provider Response Signal     & 6.57\%  \\[0.45ex]
Tool Action Payload          & 21.71\% \\[0.45ex]
Agent Context State          & 21.71\% \\[0.45ex]
Observation Feedback Signal  & 25.00\% \\
\end{tabular}

\\
\bottomrule
\end{tabular}%
}
\end{table*}

\parabf{Taxonomy of Semantic Patterns in \LDCR{} Branch Predicates.}
Table~\ref{tab:rq2_category_distribution} presents the taxonomy of semantic patterns in \LDCR{} branch predicates along with their distribution across the sample. Given the space limits, the detailed explanation and example of each category are in our artifact~\cite{harnesstester2026}. 

\textit{Component Intention.} We observe six main categories of component intention across the inspected \LDCR{} branch predicates. Among them, provider I/O normalization and tool action handling account for 29.14\% and 23.57\% of \LDCR{} branch predicates, respectively. Together, these two categories comprise more than half of the inspected predicates, showing that \LDCR{} branch predicates frequently govern the processing of provider responses and the selection, validation, and gating of model-proposed actions. This concentration highlights the important role of \LDCR{} branch predicates in mediating the transition from model-produced information to downstream agent execution.

\textit{Judgment Pattern.} Almost half of \LDCR{} branch predicates perform presence or shape checks (e.g., 47.14\%), which mainly check the structures of LLM-derived values (e.g., expected type, fields, schema, or container structure). Other recurring patterns enumerate the common properties, such as concrete values, sizes, content patterns, or policies. Overall, LLM-dependent harness branch predicates primarily determine (i) whether model-derived information is available in the expected structure, and (ii) which execution path to follow based on its value. Testing \LDCR{} therefore requires systematically varying both the structure and values of model-derived inputs to exercise distinct branch outcomes, rather than randomly constructing plausible model responses.

\textit{LLM Value Form.} We observe five forms of LLM-derived values. In addition to immediate model outputs, such as generated content and provider response signals, accounting for 25.00\% and 6.57\% of \LDCR{} branch predicates, \LDCR{} branch predicates also process (i) indirect LLM-derived values accumulated throughout agent execution, such as agent context state (21.71\%), and (ii) heterogeneous runtime values that depend on model outputs, such as tool action payloads (21.71\%) and observation feedback signals (25.00\%). These results highlight the challenge of \LDCR{} testing, as tests need to reproduce not only immediate model outputs but also heterogeneous values and states accumulated throughout long-horizon agent execution.

\finding{LLM-dependent harness branch predicates exhibit substantial semantic diversity in what they control, how they make decisions, and what forms of LLM-derived values they process. They operate not only on direct model outputs but also on heterogeneous information accumulated throughout agent execution. Adequate \LDCR{} testing is therefore challenging: tests need to reproduce different forms of LLM-derived values throughout agent execution and systematically vary their structures and properties, while ensuring that these variations satisfy  the relevant agent harness contracts.}

\begin{figure}[!hbt]
\centering
\includegraphics[width=\linewidth]{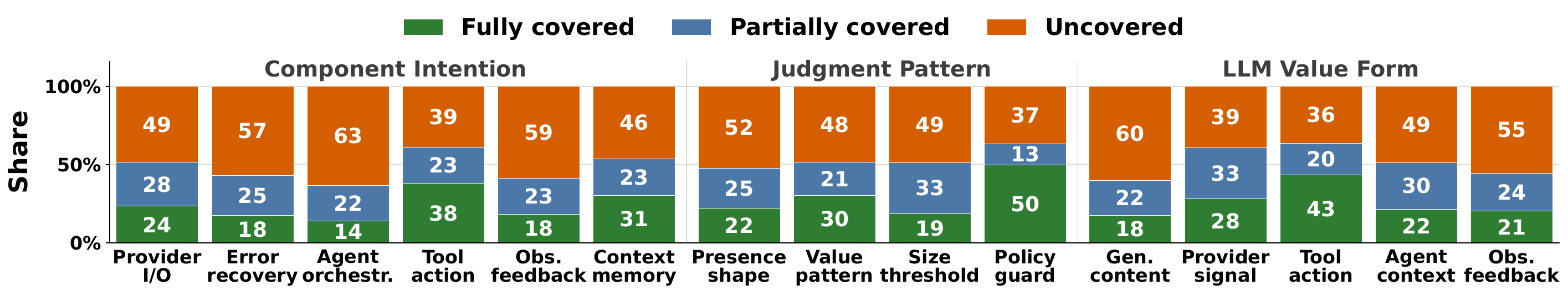}
\caption{Test coverage across \LDCR{} branch predicates.}
\label{fig:ldcr_branch_coverage}
\end{figure}

\parabf{Test Coverage Across Semantic Patterns.}
We further examine whether existing tests cover different semantic categories of \LDCR{} branch predicates equally well or leave gaps in particular categories. We classify each \LDCR{} branch predicate as \emph{fully covered}, \emph{partially covered}, or \emph{uncovered}, depending on whether all, some but not all, or none of its decision outcomes are exercised. We report the coverage by category in Figure~\ref{fig:ldcr_branch_coverage}.

Overall, among previously sampled 700 \LDCR{} branch predicates, only 25.57\% are fully covered, 24.57\% are partially covered, and 49.86\% remain entirely uncovered. This pattern separates two forms of incompleteness: many LLM-dependent predicate decisions are never reached, while many others are reached without all outcomes of the encoded decision boundary being exercised. 

\textit{Coverage Across Component Intention.} A clear coverage contrast emerges across different component intentions. Tool action handling achieves the highest full-coverage rate at 38.18\% (63/165), with 61.21\% (101/165) of its branch predicates having at least one outcome exercised. In contrast, agent orchestration control predicates exhibit lowest coverage, only 14.29\% (14/98) of predicates are fully covered, while 63.27\% (62/98) remain uncovered. Observation feedback and error recovery handling show similarly low full-coverage rates of 18.39\% (16/87) and 17.65\% (9/51), respectively. One possible explanation is that tool actions typically follow explicit schemas, making it easier to construct valid payloads and systematically vary them. In contrast, orchestration, feedback, and error recovery often depend on dynamically evolving execution contexts, including prior interactions, accumulated agent states, and cross-component behavior. Reproducing and varying such contexts while satisfying the relevant agent-harness contracts can therefore be more challenging.

\textit{Coverage Across Judgment Patterns.} Coverage also varies substantially across judgment patterns. Policy guards achieve the highest full-coverage rate at 50.00\% (15/30). This may be because policies are often implemented as functions or exposed through interfaces, thus providing a clear and controllable decision space: tests can directly mock the corresponding policy interfaces to return different decisions, making alternative branch outcomes relatively easy to exercise. 
In contrast, such direct guidance is not always available for other judgment patterns. Only 22.12\% of presence or shape checks, 18.69\% of size or threshold checks, and 30.47\% of value or pattern matching predicates are fully covered, with more than 48\% of each category remaining uncovered. For these patterns, the values inspected by a predicate may be produced through preceding parsing, or data transformations, making the required inputs less apparent. Constructing tests for different outcomes may therefore require reasoning backward from the target predicate to establish appropriate upstream inputs and states while satisfying the relevant agent-harness contracts.

\textit{Coverage Across Value Forms.}
Branch predicates over tool action payloads are the most thoroughly exercised: 43.42\% are fully covered, and 63.82\% have at least one exercised outcome. Full-coverage rates are substantially lower for generated content (17.71\%), observation feedback signals (20.57\%), and agent context state (21.71\%). Moreover, 60.00\% of generated-content predicates and 55.43\% of observation-feedback predicates remain uncovered.
One possible explanation is that tool action payloads typically have a relatively bounded and explicit value space: available actions and their argument schemas provide clear guidance for constructing valid inputs and systematically varying them. In contrast, generated content has a more open-ended value space, making it difficult to enumerate the content and structural variations needed to exercise downstream logic. Agent context and observation feedback can be even more challenging, as their values evolve throughout agent execution based on prior interactions, execution results, and runtime states. Their relevant value spaces are therefore difficult to enumerate in advance, requiring tests to reproduce and vary values along the execution flow while preserving the expected runtime context.

\finding{Test coverage varies substantially across different semantic categories of LLM-dependent harness branch predicates. Across three semantic dimensions, these predicates are better covered when test inputs and behaviors are explicitly defined and easy to control. For example, tool-action logic benefits from explicit action schemas and bounded value spaces, while policy guards can often be exercised by directly mocking policy decisions. In contrast, larger coverage gaps arise when testing requires reconstructing dynamically evolving execution contexts or exploring open-ended value spaces. In summary, existing tests are less effective at exercising \LDCR{} behavior as the required test context and input space become more implicit, dynamic, and difficult to enumerate.}

\section{\ourtool{}: Harness-Oriented Test Generation}
Our study above reveals prevalent test inadequacy in LLM-dependent harness  across real-world agentic systems, motivating the need for effective harness-oriented test generation techniques. However, as shown by our findings, extensively exercising LLM-dependent harness code is challenging: reaching target \LDCR{} requires not only test setups that satisfy the interaction contracts across harness components, but also diverse test inputs that represent different model-derived values and execution states within these valid setups. Motivated by our findings, we propose \ourtool{}, an LLM-driven test generation technique specifically designed to extensively exercise LLM-dependent harness code in agentic systems.

\begin{figure}[!hbt]
\centering
\includegraphics[width=0.8\linewidth]{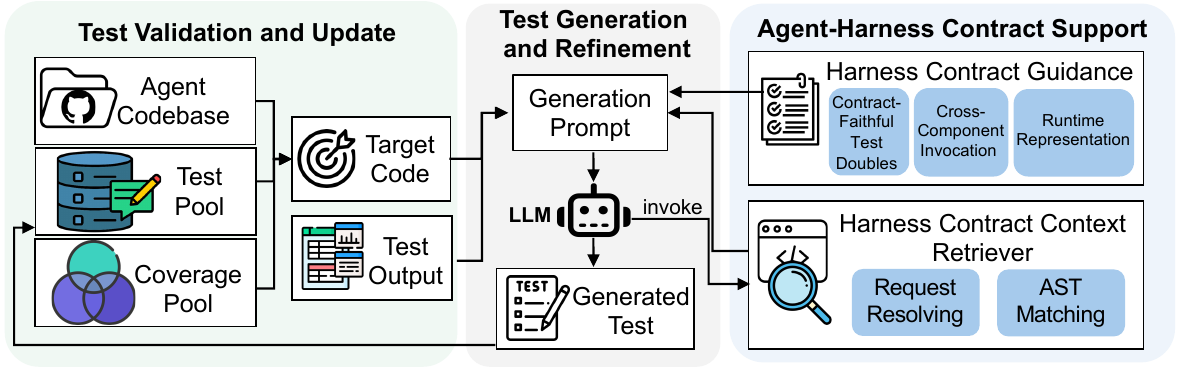}
\caption{Overview of \ourtool{}.}
\label{fig:workflow}
\end{figure}

Figure~\ref{fig:workflow} presents the overall workflow of \ourtool{}. Following the coverage-driven test generation paradigm, given an agent codebase and its existing test suite, \ourtool{} iteratively identifies \LDCR{} code that is uncovered by existing tests as targets and generates new tests to cover the target.
Different from existing LLM-driven test generation techniques designed for general software, the key insight of \ourtool{} is to make harness-oriented test generation explicitly aware of \emph{agent harness contracts}. As discussed in Section~\ref{sec:rq1}, {agent harness contracts} refer to the requirements that must hold across interactions among harness components for valid execution, such as expected data structures, execution preconditions, and interaction protocols. To this end, \ourtool{} augments general LLM-based test generation with \emph{agent harness contract support}, which helps LLM    construct contract-faithful test contexts via prompt-level guidance (i.e., agent-harness contract guidance) and tool-level enhancement (i.e., agent-harness contract context retriever). 

\textit{Agent-Harness Contract Guidance.} The guidance consists of a set of high-level, contract-aware principles embedded in the prompts used for iterative test generation and refinement. Its main goal is to guide the model to preserve key agent-harness contracts when constructing tests, including maintaining the expected runtime representations of provider responses, tool actions, and observations; constructing the agent state or interaction history required to reach the target behavior; respecting cross-component invocation contracts, such as synchronous/asynchronous calling conventions and expected return shapes; and installing contract-faithful test doubles at the actual bindings or injection points used by the code under test. These principles are derived from testing challenges identified in our study above. Beyond contract fidelity, the guidance also promotes diversity by encouraging LLM to prioritize distinct behavioral partitions rather than generating additional variants of already exercised behavior. Detailed prompts are in our artifact~\cite{harnesstester2026}. 

\textit{Agent-Harness Contract Context Retriever.} While the contract guidance above mainly provides prompt-level principles that direct LLM to important contract requirements, generating and refining a concrete harness-oriented test also requires relevant code context to show how these harness contracts are implemented in agent codebase. Therefore, to complement the prompt-level contract guidance, the harness contract context retriever further serves as a callable context collection tool, which retrieves contract-relevant code context on demand during test generation. Such context includes the code needed to construct or control the test setup, such as the definition of a constructor, fixture, collaborator, or mock target. Specifically, when invoking the retriever, the LLM requests a class, function, or method definition by providing its source file and exact symbol name, which can often be inferred from imports in the current file. For example, if the target logic requires an ``Action'' input defined elsewhere, an import such as ``\texttt{from pkg.actions import Action}'' identifies the module ``pkg.actions'', corresponding to the source file ``pkg/actions.py''. The LLM can therefore request the definition of ``Action'' from that file. The retrieved definition exposes the required fields and their types, helping the LLM construct a contract-faithful mock action that serves as a valid input for exercising the target logic. Upon receiving the request, the retriever deterministically resolves it using AST-based matching for symbol definitions and bounded extraction for surrounding module context. This demand-driven retrieval complements the high-level guidance with repository-specific context needed to construct contract-faithful test setups and make the target behavior reachable and controllable. 

\section{RQ3: Test Adequacy Improvement}
\parabf{Experiment Setting.} This RQ evaluates the effectiveness of \ourtool{} by investigating the test adequacy improvement it gains on our studied agentic systems. 

\textit{Baselines.} As there are no existing test generation techniques designed for agent harness code, we compare \ourtool{} with four state-of-the-art test generation techniques that are designed for general software systems. For the agents implemented in Python, we include \textsc{\cu{}}~\cite{pizzorno2025coverup} and \textsc{\tta{}}~\cite{zhou2026testtailor} as baselines, two recent LLM-driven test generation approaches for general Python software; for the agents implemented in TypeScript, we include \textsc{\qd{}}~\cite{qodo-cover-artifact} and \textsc{\tp{}}~\cite{schafer2024testpilot,testpilot2-artifact}, two recent LLM-driven test generation techniques for general TypeScript software. To isolate the effect of contract-specific design in \ourtool{}, we include a \textit{contract-agnostic} variant of \ourtool{} by removing the agent harness contract support. 

\textit{Experimental Process.} We include two test generation settings, including (i) \textit{general project-wide generation}, which does not specify any target but aims at the entire codebase of the studied agentic system; (ii) \textit{scoped \LDCR-wide generation}, which provides the \LDCR{}-containing files as the targets. For all baselines, we directly use their released implementation, configuring their target scope for each setting while retaining their native generation workflows. In the scoped setting, the baselines receive the same \LDCR{}-containing file as \ourtool{} for a fair comparison. We use \model{}~\cite{openai2025gpt5mini} as the backbone LLM for all studied techniques given its cost-effectiveness. 
Following previous work on application-level automated testing~\cite{yoon2024droidagent,lan2024dqt}, all techniques are executed with a 120-minute timed budget per project. 

\begin{figure}[!hbt]
    \centering
    \includegraphics[width=\linewidth]{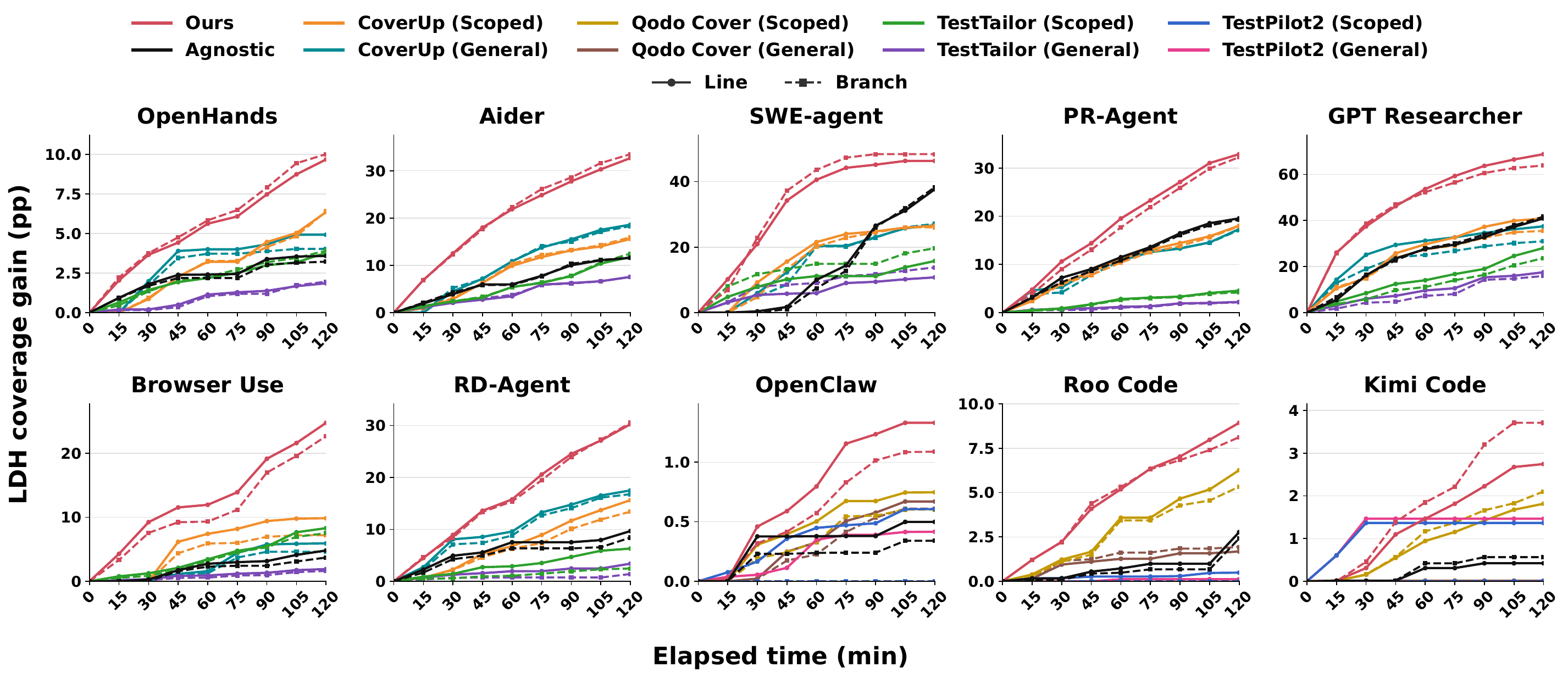}
    \caption{\LDCR{}-wide coverage gains achieved by different test generation techniques over time.}
    \label{fig:ldhc_coverage_gain}
\end{figure}

\begin{figure}[!hbt]
    \centering
    \includegraphics[width=\linewidth]{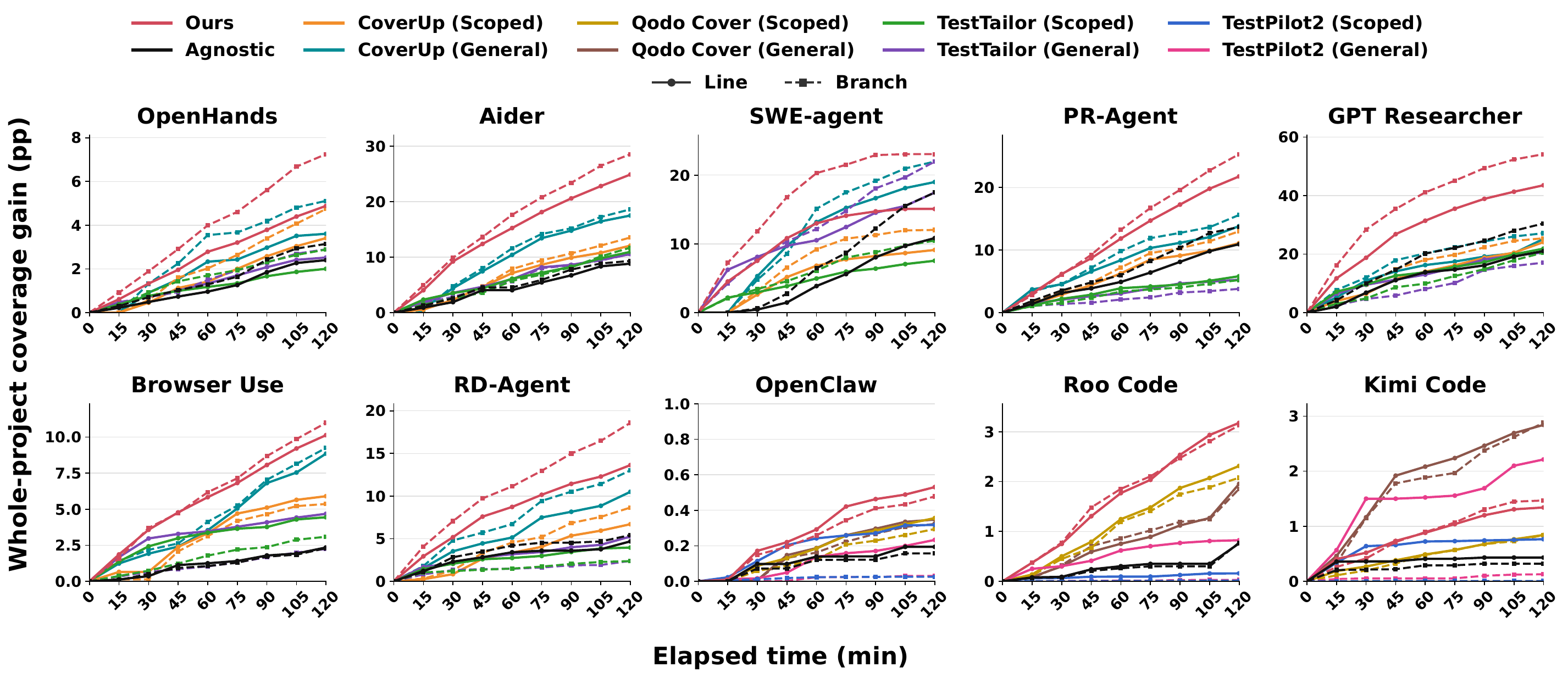}
    \caption{Project-wide coverage gains achieved by different test generation techniques over time.}
    \label{fig:project_coverage_gain}
\end{figure}

\parabf{Test Coverage Improvement.} Figure~\ref{fig:ldhc_coverage_gain} and Figure~\ref{fig:project_coverage_gain} present the trend of \LDCR{}-level and project-level coverage gains achieved by all studied techniques over time.  

\textit{Consistent \LDCR{} Coverage Gains.}  \ourtool{} consistently achieves the highest final \LDCR{}-wide line- and branch coverage gains among  all studied baselines, indicating its effectiveness in improving test adequacy for LLM-dependent harness code. Moreover, this advantage emerges early in the test-generation process: e.g., \LDCR{}-wide branch coverage gains lead across all ten projects from the 30-minute checkpoint onward, and line coverage gains do so from 75 minutes onward.

\textit{Substantial Gains in Targeted Test Generation.} Although restricting test generation to \LDCR{}-bearing files substantially benefits baselines (e.g.,  25.09\% and 23.15\% higher gains compared with the general setting), \ourtool{} still substantially outperforms all the baselines with the scoped setting. 
At 120 minutes, \ourtool{} achieves macro-averaged \LDCR{} line- and branch-coverage gains of 25.84 and 25.41 percentage points, respectively, with the line-coverage gain reaching 68.77 points on GPT Researcher. Compared with the strongest scoped baseline selected separately for each agent and metric, these average gains are 82.27\% and 94.36\% larger, respectively. Against the strongest general external baselines, the corresponding improvements are 95.29\% and 109.22\%. These results confirm the advantage of \ourtool{} over baselines in the targeted test generation scenario.

\textit{Contribution from Harness Contract Support.} The pronounced performance drop from \ourtool{} to its contract-agnostic variant (the ``Agnostic'' in Figure~\ref{fig:ldhc_coverage_gain}) confirms the effectiveness of our key design on harness contract support. In particular, \ourtool{} produces 96.51\% larger mean line-coverage gains and 96.24\% larger branch-coverage gains, outperforming the contract-agnostic variant on both metrics in every project. We further run a fine-grained ablation analysis on OpenHands using variants without harness contract guidance and the harness contract context retriever, respectively. We observe consistent decreases in \LDCR{} line/branch coverage gains (43.26\%/45.98\% and 37.38\%/41.00\%, respectively) compared to \ourtool{}, indicating the complementary contribution of each mechanism to harness contract support.

\begin{figure}[!hbt]
    \centering
    \includegraphics[width=\linewidth]{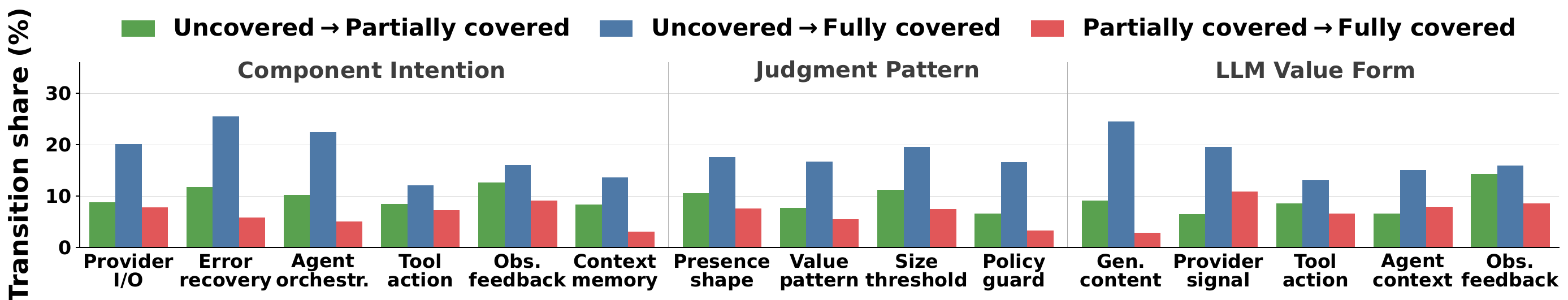}
    \caption{Semantic coverage gains achieved by \ourtool{}.}
    \label{fig:ldh_branch_cov_augment}
\end{figure}

\textit{Prevalent Improvements Across Semantic Categories.} To examine whether \ourtool{} improves test adequacy across the diverse semantic categories identified in RQ2, we further report category-level coverage improvements in Figure~\ref{fig:ldh_branch_cov_augment}. Improvements are observed across all semantic categories: 12.12--25.49\% of predicates in each category transition from uncovered to fully covered, while another 2.86--10.87\% transition from partially to fully covered. Notably, for the challenging category of \emph{Agent Orchestration Control}, where stateful and cross-component logic requires carefully coordinated test setups, 22.45\% of branch predicates move from uncovered to fully covered and another 5.10\% from partially to fully covered. These broad improvements suggest that \ourtool{} can effectively exercise semantically diverse \LDCR{} behaviors, including those requiring complex test setups. The improvement aligns with the design intuition of using contract guidance and retrieved harness context to construct contract-faithful setups with diverse inputs.

\textit{Consistent Gains in Project-wide Coverage.} As shown in Figure~\ref{fig:project_coverage_gain}, besides \LDCR{}-wide coverage, \ourtool{} also achieves the highest project-level line- and branch-coverage gains among all configurations on eight and nine of the ten agents, respectively. Despite targeting \LDCR{}, its macro-averaged gains reach 13.90 and 17.30 percentage points, exceeding the corresponding 10.36 and 11.58 points achieved by the strongest general baselines (\cu{} and \qd{}). These results provide a complementary view of \ourtool{}'s effectiveness in generating high-quality tests for agentic systems, with coverage benefits evident not only within LLM-dependent harness code but also within the whole project.

\begin{figure}[!hbt]
    \centering
    \includegraphics[width=\linewidth]{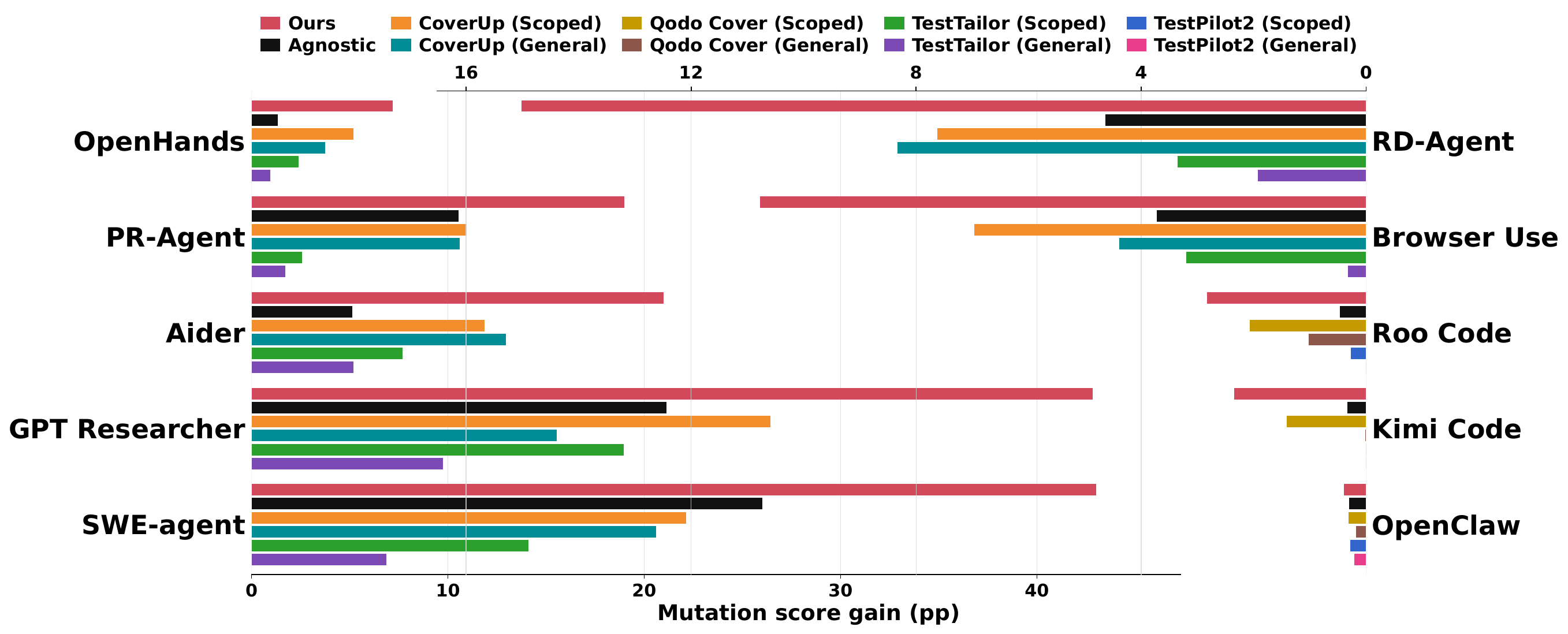}
    \caption{\LDCR{} mutation score gains achieved by different test generation techniques.}
    \label{fig:mutation_score_gain}
\end{figure}

\parabf{Mutant Detection Effectiveness.}
Figure~\ref{fig:mutation_score_gain} presents the mutation score gains achieved by all studied techniques. Overall, \ourtool{} achieves the highest mutation-score gain on every studied agent, outperforming all baselines and its contract-agnostic variant. Its largest percentage-point margin over the strongest baseline occurs on SWE-agent, where it gains 43.01 points compared with 22.15 for \cu{} (Scoped). Its advantage on OpenClaw appears smaller in percentage-point terms, partly resulting from the large measurement denominator: \ourtool{}'s 0.39-point gain corresponds to 185 additional kills over a pool of 46,967 mutants, compared with 147 for \qd{} (Scoped) and 141 for the contract-agnostic variant. Despite the small percentage-point differences, \ourtool{} nevertheless kills 38 and 44 more mutants, respectively. Manual inspection of the generated tests and mutation results shows that they kill both mutants at previously uncovered locations and mutants that survived despite being exercised by the original suite. Overall, the improved capabilities of mutant detection further support \ourtool{}'s effectiveness in improving agent-harness test adequacy.

\finding{\ourtool{} demonstrates strong capability in generating high-adequacy tests for LLM-dependent harness code across all studied agents. It consistently outperforms all evaluated test generation baselines in both \LDCR{}- and project-wide coverage gains. Our ablation study further demonstrates the substantial contribution of agent-harness contract support. Beyond coverage, tests generated by \ourtool{} also exhibit stronger mutant detection capability. Overall, these results highlight the value of developing harness-oriented test generation techniques to improve the test adequacy of real-world agentic systems.}

\section{RQ4: Real-World Harness Bug Detection}
Although mutation scores can reflect the potential fault-detection capability of generated tests, mutants are artificially injected faults and may not fully represent real-world harness bugs. Therefore, in this RQ, we further evaluate the test adequacy of tests generated by \ourtool{} and the baselines by examining their capability to detect real-world harness bugs.

\subsection{RQ 4.a: Detecting Historical Harness Bugs}
\parabf{Benchmark and Process.} To systematically evaluate the precision and recall of bug detection, we first build  \ourbench{}, a benchmark of 250 real-world harness bugs historically reported and fixed across 10 studied agentic systems, including 140 bugs from seven Python agents and 110 from three TypeScript agents. We construct the benchmark by mining historical bug-fixing commits and retaining bugs whose patches involve \LDCR{} code, thereby focusing on bugs in LLM-dependent harness logic, given its importance~\cite{chen2026agentreactive}. To reduce leakage from existing regression tests, we retain only bugs that cannot be exposed by any test in the buggy revision, preventing the LLMs from simply reusing existing bug-revealing tests for test generation. To the best of our knowledge, \ourbench{} is the first benchmark dedicated to real-world LLM-dependent harness bugs, and we release it in our artifact~\cite{harnesstester2026} for future research.

Given the large manual effort involved in inspecting test failures, in this RQ, we compare \ourtool{} with the two best-performing baselines in RQ3 (i.e., \textsc{\cu{}} for Python and \textsc{\qd{}} for TypeScript). To assess the contribution of harness contract support, we also include the contract-agnostic variant of \ourtool{}. In particular, for each case, we provide each technique with the buggy location (e.g., the functions or classes where the developer-submitted patch occurs) as the target code for test generation, but without exposing the patch, bug description, or fixed-side implementation. A bug is considered plausibly detected only if the generated test fails on the buggy version but passes on the fixed version. We then manually review all plausible detections to verify whether the test exposes the same bug fixed by developers or instead constitutes a false positive.

\setlength{\intextsep}{0pt}
\setlength{\histtablewidth}{0.50\textwidth}
\begin{wraptable}{r}[0pt]{\histtablewidth}
\centering
\captionsetup{position=top,skip=3pt}
\caption{Historical bug revealing.}
\label{tab:historical-bug-revealing}

\small
\setlength{\tabcolsep}{2.5pt}
\renewcommand{\arraystretch}{1.05}

\resizebox{\histtablewidth}{!}{%
\begin{tabular}{@{}lrrrrrrr@{}}
\toprule
& & \multicolumn{2}{c}{\makebox[0pt][c]{\textbf{\ourtool{}}}}
  & \multicolumn{2}{c}{\textbf{Agnostic}}
  & \multicolumn{2}{c}{\textbf{Baseline}} \\
\cmidrule(lr){3-4}
\cmidrule(lr){5-6}
\cmidrule(lr){7-8}
\textbf{Agent} & \textbf{Cases}
& \textbf{TP/FP} & \shortstack{\textbf{Cost\$}}
& \textbf{TP/FP} & \shortstack{\textbf{Cost\$}}
& \textbf{TP/FP} & \shortstack{\textbf{Cost\$}} \\
\midrule
OpenHands      & 42 & 8/0  & 15.00 & 2/1 & 9.86  & 0/0 & 3.91 \\
Aider          & 19 & 6/1  & 5.19  & 1/1 & 4.51  & 0/0 & 1.35 \\
SWE-agent      & 7  & 2/0  & 1.98  & 1/0 & 1.40  & 0/0 & 0.71 \\
PR-Agent       & 11 & 2/1  & 3.28  & 1/0 & 3.03  & 0/0 & 2.21 \\
GPT Researcher & 4  & 3/1  & 0.66  & 3/0 & 0.38  & 0/0 & 0.13 \\
Browser Use    & 41 & 8/2  & 13.91 & 3/0 & 15.56 & 0/0 & 3.85 \\
RD-Agent       & 16 & 1/0  & 6.77  & 0/0 & 4.47  & 0/0 & 1.13 \\
OpenClaw       & 88 & 10/1 & 22.72 & 4/3 & 19.54 & 2/0 & 6.23 \\
Roo Code       & 16 & 2/0  & 6.17  & 0/0 & 4.58  & 0/0 & 1.42 \\
Kimi Code      & 6  & 1/1  & 1.89  & 0/0 & 1.59  & 0/0 & 0.75 \\
\midrule
\textbf{Total} & \textbf{250}
& \textbf{43/7} & \textbf{77.58}
& \textbf{15/5} & \textbf{64.92}
& \textbf{2/0}  & \textbf{21.68} \\
\bottomrule
\end{tabular}%
}
\end{wraptable}

\parabf{Results.} Table~\ref{tab:historical-bug-revealing} presents the number of detected true bugs (TP) and false positives (FP) of all studied techniques on \ourbench{} with the total generation costs per project. Overall, \ourtool{} achieves the highest bug detection rate (17.20\%) by detecting 43 historical bugs, while the best baseline can only detect 2 historical bugs (0.80\%). In addition, removing harness-contract support reduces the number of revealed bugs from 43 (17.20\%) to 15 (6.00\%), while the proportion of false positives among retained differential outcomes increases from 14.00\% to 25.00\%. Manual review identified typical false positives arising from fixtures that violated production-side agent-harness contracts or assertions that assumed unsupported behavior. For example, one test omitted a required command field from a tool-call payload but expected an undocumented fallback; another passed a bracketed IP literal to a helper whose caller supplies normalized hostnames. These cases reinforce the need for agent-harness contract fidelity: fixtures and mocks must preserve production-established contracts to help the test reach the target in a valid way, and must not demand behavior beyond existing contracts.
In terms of cost, \ourtool{} is more expensive overall but substantially more cost-effective for bug revealing. Across 250 cases, \ourtool{} reveals 43 bugs at a total cost of \$77.58, or \$1.80 per revealed bug, compared with \$10.84 per bug for the baseline.

\subsection{RQ 4.b: Detecting Previously-Unknown Harness Bugs} 
To further investigate whether \ourtool{} can generate tests to detect previously-unknown harness bugs, we apply it to the latest releases of agentic systems used in our previous study. In particular, as OpenHands was recently refactored into several components, we therefore select its two core components, Agent Canvas~\cite{openhands-agent-canvas} and SDK~\cite{openhands-software-agent-sdk} in this experiment. To control costs, we select 100 functions containing \LDCR{} code per agent codebase as test-generation targets. Then, we manually inspect all failures triggered by the generated tests and report new bugs to developers. 
\par
\setlength{\intextsep}{0pt}
\setlength{\bugtablewidth}{0.5\textwidth}

\begin{wraptable}{r}[0pt]{\bugtablewidth}
\centering
\captionsetup{position=top,skip=3pt}
\caption{Detected real-world bugs.}
\label{tab:latest-revision-bugs}

\small
\setlength{\tabcolsep}{2.5pt}
\renewcommand{\arraystretch}{1.05}

\resizebox{\bugtablewidth}{!}{%
\begin{tabular}{@{}lrrrrr@{}}
\toprule
\textbf{Project}
& \textbf{New}
& \textbf{Known}
& \textbf{Confirmed}
& \shortstack[r]{\textbf{Fix}}
& \shortstack{\textbf{Cost\$}} \\
\midrule
OpenHands Canvas & 5  & 0 & 3  & 3  & 7.38 \\
OpenHands SDK    & 10 & 3 & 10 & 9  & 7.68 \\
Aider            & 9  & 4 & 9  & 9  & 6.69 \\
SWE-agent        & 4  & 3 & 4  & 4  & 6.71 \\
PR-Agent         & 5  & 2 & 5  & 5  & 9.16 \\
GPT Researcher   & 6  & 1 & 6  & 6  & 5.80 \\
Browser Use      & 15 & 8 & 13 & 12 & 7.85 \\
RD-Agent         & 11 & 5 & 9  & 8  & 8.23 \\
OpenClaw         & 9  & 6 & 9  & 8  & 7.15 \\
Roo Code         & 11 & 0 & -- & -- & 7.26 \\
Kimi Code        & 3  & 2 & 1  & 0  & 7.73 \\
\midrule
\textbf{Total}
& \textbf{88}
& \textbf{34}
& \textbf{69}
& \textbf{64}
& \textbf{81.64} \\
\bottomrule
\end{tabular}%
}
\end{wraptable}

\parabf{Results.} Table~\ref{tab:latest-revision-bugs} summarizes our detected bugs and their upstream outcomes. In total, \ourtool{} identifies 122 true harness bugs, including 34 previously-known and 88 previously-unknown bugs. Among the 88 previously unknown bugs, 69 are confirmed by developers, 64 are fixed or assigned an open fix, and one of them is assigned a CVE number. Roo Code's GitHub repository was archived for refactoring during our study period; therefore, new issues could not be created. Overall, these results highlight the practical effectiveness of \ourtool{} in generating bug-revealing tests for agent harnesses. 

\parabf{Real-World Example.}
Listing~\ref{lst:nested-secret} shows a previously-unknown sensitive-data leakage bug in Browser Use detected by \ourtool{} and confirmed by developers. The system records model-selected tool calls as typed actions in the agent execution history. Our generated test reveals that a sensitive value nested within a typed action parameter can bypass redaction and remain in the saved history. In the persistence path, \texttt{dump\_actions} serializes the actions, while \texttt{redact\_dict} delegates list-valued fields to \texttt{redact\_list}, which handles strings and dictionaries but leaves nested lists unchanged. Consequently, the synthetic credential nested in \texttt{rows} remains unredacted in the saved history. Revealing this bug requires more than placing a secret in an arbitrary payload. The test must embed a schema-valid input in a valid chain of action, model-output, and history objects, provide the expected sensitive-data mapping, and inspect the resulting file. This illustrates the role of agent-harness contracts: the test setup must preserve the runtime representations, object relationships, and interactions expected by the target logic. The test must further vary this valid setup with a nested-list input containing the sensitive value. Neither the best-performing Python baseline (\cu{}) nor our contract-agnostic variant revealed this bug. After confirmation, the bug was fixed upstream through recursive redaction.

\begin{figure}[!hbt]
\centering
\refstepcounter{lstlisting}
\label{lst:nested-secret}
\captionsetup{position=top,skip=6pt}
\caption*{\textbf{Listing~\thelstlisting.} A previously-unknown sensitive-data leakage in Browser Use detected by \ourtool{}.}

\newcommand{\entryhl}[1]{%
  \begingroup
  \setlength{\fboxsep}{0.4pt}%
  \colorbox{yellow!38}{\ttfamily\scriptsize #1}%
  \endgroup
}
\newcommand{\bughl}[1]{%
  \begingroup
  \setlength{\fboxsep}{0.4pt}%
  \colorbox{cyan!26}{\ttfamily\scriptsize #1}%
  \endgroup
}
\newcommand{\pathhl}[1]{%
  \begingroup
  \setlength{\fboxsep}{0.4pt}%
  \colorbox{cyan!24}{\ttfamily\scriptsize #1}%
  \endgroup
}
\newcommand{\setuphl}[1]{%
  \begingroup
  \setlength{\fboxsep}{0.4pt}%
  \colorbox{yellow!28}{\ttfamily\scriptsize #1}%
  \endgroup
}
\newcommand{\triggerhl}[1]{%
  \begingroup
  \setlength{\fboxsep}{0.4pt}%
  \colorbox{green!28}{\ttfamily\scriptsize #1}%
  \endgroup
}
\newcommand{\oraclehl}[1]{%
  \begingroup
  \setlength{\fboxsep}{0.4pt}%
  \colorbox{red!18}{\ttfamily\scriptsize #1}%
  \endgroup
}

\lstset{
  style=pyflowpanel,
  escapeinside={(*@}{@*)}
}

\begin{minipage}[t]{0.50\linewidth}
\vspace{0pt}
\begin{lstlisting}
# Simplified persistence path
class AgentHistoryList:
    def (*@\entryhl{save\_to\_file}@*)(self, out, sensitive_data):
        data = {"actions": dump_actions(self.history)}
        write_json(out,
                   (*@\pathhl{redact\_dict}@*)(data, sensitive_data))

def (*@\pathhl{redact\_dict}@*)(data, s):
    return {
        k: mask(v, s) if isinstance(v, str)
        else redact_dict(v, s) if isinstance(v, dict)
        else (*@\bughl{redact\_list(v, s)}@*) if isinstance(v, list)
        else v for k, v in data.items()
    }

# Bug: nested lists remain unchanged
def (*@\bughl{redact\_list}@*)(value, s):
    return [
        mask(x, s) if isinstance(x, str)
        else redact_dict(x, s) if isinstance(x, dict)
        (*@\bughl{else x}@*) for x in value
    ]
\end{lstlisting}
\end{minipage}\hfill%
\begin{minipage}[t]{0.50\linewidth}
\vspace{0pt}
\begin{lstlisting}
# Test body; imports and wrapper omitted
class InputParams(BaseModel):
    (*@\entryhl{rows: list[list[str]]}@*)
class InputAction(ActionModel):
    input: InputParams

secret = "token-123"
custom_output = (*@\entryhl{AgentOutput.type\_with\_custom\_actions}@*)(
    InputAction)
history = (*@\entryhl{AgentHistoryList(history=[AgentHistory(}@*)
    model_output=custom_output(
        evaluation_previous_goal="ok", memory="ok",
        next_goal="done", action=[InputAction(
            input=InputParams((*@\triggerhl{rows=[[secret]]}@*)))]),
    result=[], (*@\entryhl{state=BrowserStateHistory(}@*)
        url="https://example.com", title="Example",
        tabs=[], interacted_element=[]))])
out = tmp_path / "history.json"
history.(*@\entryhl{save\_to\_file}@*)(
    out, (*@\entryhl{sensitive\_data=\{"api\_key": secret\}}@*))
saved = out.read_text(encoding="utf-8")
(*@\oraclehl{assert secret not in saved}@*)  # fails
(*@\oraclehl{assert "<secret>api\_key</secret>" in saved}@*)
\end{lstlisting}
\end{minipage}
\end{figure}

\finding{\ourtool{} effectively detects real-world harness bugs in agentic systems. In particular, tests generated by \ourtool{} can detect substantially more historical bugs than the best-performing baseline. Beyond historical bugs, \ourtool{} also uncovers 88 previously unknown harness bugs, 69 of which have been confirmed by developers. These results demonstrate the practical effectiveness of \ourtool{} in uncovering harness bugs and improving the reliability of agentic systems.}

\section{Threats to Validity}
\textit{External Validity.}
The generality of our findings can be scoped to our studied agents. To mitigate this threat, we select diverse and actively developed systems (e.g., OpenClaw) spanning different application domains, implementation languages, and agent architectures. Similarly, our evaluation considers a representative set of existing test generation techniques and configurations, while other techniques or models may yield different results. We therefore release our benchmark and implementation to facilitate replication and evaluation on additional systems and techniques~\cite{harnesstester2026}. 
\textit{Internal Validity.} The semantic pattern analysis in RQ2 relies on manual labeling and may be subject to human judgment. To reduce this threat, two experienced annotators independently label the sampled branches using an explicitly defined codebook, achieving high inter-annotator agreement ($\kappa=0.96$), with remaining disagreements resolved through discussion with a third reviewer. Our metric calculation and implementation may be subject to implementation errors. To mitigate this threat, we validate intermediate results, manually inspect representative cases, and use established coverage and mutation-testing tools whenever applicable to reduce potential measurement errors.

\section{Conclusion}
\label{sec:conclusion}
This work presents the first study of test adequacy in real-world agent harness and reveals substantial test-adequacy gaps in exercising the critical LLM-dependent harness code. We find that these gaps are particularly pronounced when testing requires dynamic execution contexts, implicit input spaces, and cross-component interactions. Motivated by these findings, we propose \ourtool{}, the first harness-oriented test generation technique that incorporates agent-harness contract support to construct contract-faithful tests for LLM-dependent harness code. Extensive evaluation demonstrates that \ourtool{} substantially outperforms general-purpose test generation in coverage and mutation-score gains, while its contract support contributes significantly to these improvements. Furthermore, \ourtool{} detects 122 real-world harness bugs in widely-used agentic systems (e.g., OpenClaw). 88 of them are previously-unknown bugs and 69 of them have been confirmed by developers. Overall, our findings highlight the practical value of building harness-oriented test generation in improving the reliability of real-world agentic systems.

\section*{Data Availability}
The data and code of this work are available at ~\cite{harnesstester2026}. 

\bibliographystyle{ACM-Reference-Format}
\bibliography{reference}

\end{document}